\documentclass[iop, twocolumn]{aastex6}
\usepackage{amsmath}     
\usepackage{wasysym}           
\usepackage{graphicx}
\usepackage{amssymb}
\usepackage{epstopdf}
\usepackage{mathrsfs}
\usepackage{natbib}
\usepackage{color}
\usepackage{lipsum}

\begin{document}
\title{Prompt planetesimal formation beyond the snow line}
\author{Philip J. Armitage\altaffilmark{1,2}, Josh A. Eisner\altaffilmark{3,1}, and Jacob B. Simon\altaffilmark{4,5}}

\altaffiltext{1}{JILA, University of Colorado and NIST, 440 UCB, Boulder, CO 80309-0440}
\altaffiltext{2}{Department of Astrophysical and Planetary Sciences, University of Colorado, Boulder}
\altaffiltext{3}{Steward Observatory, University of Arizona, 933 N. Cherry Ave, Tucson, AZ 85721-0065}
\altaffiltext{4}{Department of Space Studies, Southwest Research Institute, Boulder, CO 80302}
\altaffiltext{5}{Sagan Fellow}

\email{pja@jilau1.colorado.edu}

\begin{abstract}
We develop a simple model to predict the radial distribution of planetesimal formation. 
The model is based on the observed growth of dust to mm-sized particles, 
which drift radially, pile-up, and form planetesimals where the stopping time and dust-to-gas 
ratio intersect the allowed region for streaming instability-induced gravitational collapse. Using an 
approximate analytic treatment, we first show that drifting particles define a track in metallicity--stopping 
time space whose only substantial dependence is on the disk's angular momentum transport efficiency. 
Prompt planetesimal formation is feasible for high particle accretion rates (relative to the gas, 
$\dot{M}_p / \dot{M} \gtrsim 3 \times 10^{-2}$ for $\alpha = 10^{-2}$), that could only 
be sustained for a limited period of time. If it is possible, it would lead to the 
deposition of a broad and massive belt of planetesimals with a sharp outer edge. Including turbulent 
diffusion and vapor condensation processes numerically, we find that a modest enhancement 
of solids near the snow line occurs for cm-sized particles, but that this is largely 
immaterial for planetesimal formation. We note that radial drift couples planetesimal formation 
across radii in the disk, and suggest that considerations of planetesimal formation favor 
a model in which the initial deposition of material for giant planet cores occurs well beyond the snow line.
\end{abstract}


\section{Introduction}
Where and when planetesimals form within protoplanetary disks set the initial conditions for 
the gravity-dominated phase of planet formation. Meteoritic evidence is consistent 
with planetesimals forming early, with \cite{schiller15} suggesting that differentiated bodies 
formed only 0.25~Myr after calcium-aluminum-rich inclusions, but little is known empirically about the radial 
profile beyond the general observation that the Solar System, and debris disks,  
formed planetesimals at a range of radii. One can try to back out the initial distribution 
of planetesimals from the observed architecture of planetary systems \citep[as in the 
Minimum Mass Solar Nebula,][]{hayashi81}, but this is an ill-posed problem if 
planets migrate \citep{kley12}. Given recent 
advances in the characterization of gas and dust in protoplanetary 
disks, a forward-modeling approach that combines observations with 
planetesimal formation theory may prove at least as instructive.

Here, we develop a simplified global model for where planetesimals form based on 
known physical processes. Experiments and related modeling suggest 
that dust grows rapidly up to $\sim$mm sizes, defined by the onset of 
bouncing \citep{blum08,zsom10}. We assume that this remains 
true beyond the snow line, although the physics of ice 
coagulation can differ substantially from that of silicates \citep{dominik97,gundlach15}, 
in a direction that allows larger particles to form \citep{wada11,kataoka13}.
The small macroscopic solids that result from coagulation are 
then subject to radial drift \citep{weidenschilling77} and turbulent diffusion \citep{clarke88}, 
which are sufficiently rapid as to lead to an approximate steady-state on small 
scales. As the ice-dominated solids cross the snow line they evaporate, and the vapor 
diffuses outward and recondenses forming an enhancement of the solid surface 
density \citep{stevenson88}. At radii where the dust-to-gas ratio and dimensionless 
stopping time fall within certain ranges \citep[determined by][]{carrera15}, 
the streaming instability \citep{youdin05} results in the rapid formation 
of planetesimals. Planetesimals formed from the streaming instability are characteristically 
large \citep{johansen07,johansen12,simon16}, and to a good approximation they will be 
immune to aerodynamic drift and stay in place.

The above sketch defines a model for ``prompt" 
planetesimal formation that does not invoke helpful but less 
well-understood processes \citep[such as ``lucky growth" beyond material barriers, 
and local concentration in zonal flows, vortices or persistent particle 
traps, for reviews see][]{johansen14,armitage15}. It is obviously incomplete. 
Our intent is to highlight in the simplest possible model the critical role of the competition between 
radial drift and planetesimal formation.

\section{A model for planetesimal formation from drifting solids}
\label{sec_analytic}
Consider a gas disk around a star of mass $M_*$ that is parameterized by the accretion 
rate $\dot{M}$, angular momentum transport efficiency parameter $\alpha$ \citep{shakura73}, 
and central temperature profile $T(r) \propto r^{-\beta}$, with $\beta$ a constant. In steady-state,
\begin{equation}
 \nu \Sigma = \frac{\dot{M}}{3 \pi},
\end{equation}
where the effective viscosity is given in terms of the sound speed $c_s$ and angular velocity 
$\Omega$ by $\nu = \alpha c_s^2 / \Omega$, and $c_s^2 = k_{\rm B} T / \mu m_H$. Here 
$k_B$ is the Boltzmann constant, $\mu \simeq 2.2$ is the mean molecular weight, and $m_H$ 
is the mass of a hydrogen atom. The central density of the disk is $\rho_0 = \Sigma / \sqrt{2 \pi} h$, 
where $\Sigma$ is the surface density and $h = c_s / \Omega$ is the vertical scale height.

A constant $\alpha$ model provides a convenient link between $\dot{M}$ and $\Sigma$ using 
only a single free parameter, though it is unlikely to provide a good representation of 
protoplanetary disks. Simulations show that the effective $\alpha$ due to magnetohydrodynamic processes 
varies with radius \citep{simon15}, and that the accretion stress is largely divorced from the 
strength of turbulent diffusion. (In what follows we implicitly assume that turbulent diffusion 
is weak, allowing particles to settle.) For particle flows, however, what matters most are the profiles of surface density, 
temperature, and (particularly) pressure. Our fiducial model has $T = 150 (r / 3 \ {\rm AU})^{-0.5} \ {\rm K}$ 
and $\Sigma \propto r^{-1}$. The normalized pressure gradient parameter,
\begin{equation}
 \Delta \equiv - \frac{1}{2} \left( \frac{c_s}{v_K} \right) \frac{{\rm d} \ln P}{{\rm d} \ln r},
\end{equation} 
which determines in part the strength of the streaming instability \citep{bai10}, 
has a value ($\Delta \simeq 0.05$ at 1~AU) that is compatible with the simulations we 
compare against \citep{carrera15}.

\begin{figure}
\begin{center}
\includegraphics[width=0.45\textwidth,angle=0]{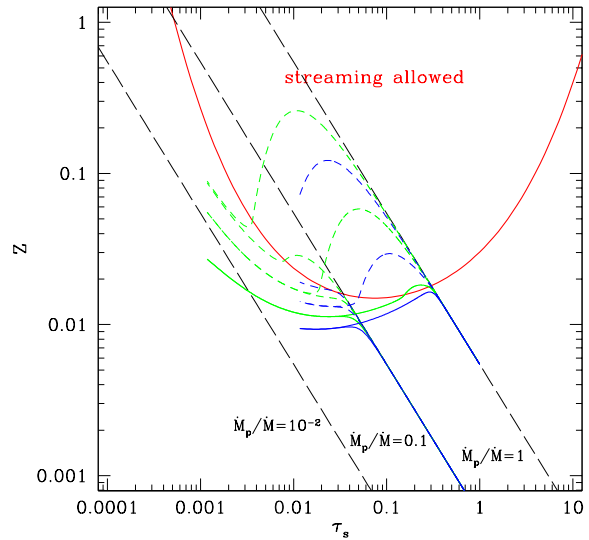}
\end{center}
\vspace{-0.3truein}
\caption{Analytic tracks (dashed lines) in $Z$-$\tau_s$ space for a disk with $\alpha = 10^{-2}$, $\beta = 1/2$, 
and various ratios of the particle to gas accretion rate. Particles evolve from lower right to upper left under 
radial drift. The red curve defines the approximate region within 
which prompt planetesimal formation is possible. The green (for mm-sized particles) and blue (cm-sized) curves 
show tracks that include a sink term for planetesimal formation, for $K = 10^2$ (solid curves), $K=10^3$ and $K=10^4$ (dashed  
curves). The disk model has $\dot{M} = 10^{-8} \ M_\odot \ {\rm yr}^{-1}$ and $T = 150 (r / 3 \ {\rm AU})^{-0.5} \ {\rm K}$.}
\label{fig1}
\end{figure}

To develop an approximate analytic model for the distribution of solids within 
the disk we note that small macroscopic particles, that have grown by coagulation outside the snow line, 
fall in the Epstein drag regime. For particles of material density $\rho_{\rm m}$ and 
radius $s$, the dimensionless stopping time is,
\begin{equation}
 \tau_s = \frac{\pi}{2}\frac{\rho_{\rm m}}{\Sigma} s.
\end{equation}
The aerodynamic drift rate is \citep{takeuchi02},
\begin{equation}
 v_{\rm r} = \frac{\tau_s^{-1} v_{\rm r, gas} - \eta v_K}{\tau_s + \tau_s^{-1}},
\end{equation} 
where $v_{\rm r, gas}$ is the radial velocity of the gas, $v_K$ is the Keplerian 
velocity, and $\eta = - (h/r)^2 (\beta / 2 - 3)$ is a parameter measuring the 
degree of radial pressure support for the gas disk. We take the limit where 
$\tau_s$ is small and $| v_{\rm r,gas} | \ll | \eta \tau_s v_K |$. For a steady-state 
radial particle flow, with accretion rate $\dot{M}_p$ and surface density $\Sigma_p$, 
\begin{equation}
 \frac{\Sigma_p}{\Sigma} = \frac{3 \alpha}{2 (3 - \beta / 2)} \frac{\dot{M}_p}{\dot{M}} \tau_s^{-1}.
\label{eq_track} 
\end{equation} 
This expression shows how the solid-phase ``metallicity" $Z \equiv \Sigma_p / \Sigma$ 
depends upon the disk properties and ratio of solid to gas accretion rates, for given $\tau_s$.

Writing the familiar physics of radial drift in the above form is useful because  
the conditions for the streaming instability to lead to gravitational collapse and  
planetesimals can {\em also} be expressed as an allowed region in $Z$-$\tau_s$ space.  
A polynomial fit to the simulation results of \cite{carrera15} gives the critical 
metallicity as\footnote{\cite{carrera15} model disks with negligibly weak 
intrinsic turbulence. Significant levels of turbulence would impede settling and planetesimal 
formation, especially at low $\tau_s$. Our results are thus most applicable to disks in which 
the accretion stress is primarily laminar, with low levels of turbulence and attendant diffusion.},
\begin{eqnarray}
 \log Z_{\rm crit} = 2.84 \times 10^{-2} (\log \tau_s)^4 + 0.11 (\log \tau_s)^3 \nonumber \\ 
 + 0.38 (\log \tau_s)^2 + 0.6 (\log \tau_s) -1.52.
\end{eqnarray} 
Equation~(\ref{eq_track}) then defines a line in that space that either does or does not 
admit planetesimal formation. Whether planetesimal formation is possible depends on $\alpha$ and the particle flux 
but not on the actual physical size to which solids grow. 
Plotting these tracks in Figure~\ref{fig1} for $\alpha = 10^{-2}$ \citep[at the high end of 
observationally estimated values,][]{hartmann98}, we see that if $\dot{M}_p / \dot{M} = 10^{-2}$ 
the metallicity is too low at any $\tau_s$ to allow prompt 
planetesimal formation. Planetesimal formation is only possible in the outer disk 
if $\dot{M}_p / \dot{M} \gtrsim 3 \times 10^{-2}$, i.e. for relative accretion rates that 
exceed the fiducial dust-to-gas ratio of $10^{-2}$. This implies that planetesimal 
formation would occur while the global average of the dust-to-gas ratio was decreasing 
with time, presumably early on. 

A steady-state model will only be valid, even approximately, if particles can grow to a size 
set by material barriers before they drift significantly \citep[otherwise they will be in the regime 
of ``drift-limited growth",][]{birnstiel12}. \cite{birnstiel12} estimate that a particle grows from 
radius $s_0$ to $s$ on a time scale,
\begin{equation}
 t_{\rm grow} \approx \frac{1}{\Omega} \frac{\Sigma}{\Sigma_p} \ln \left( \frac{s}{s_0} \right).
\end{equation}
Adopting $s_0 = \mu {\rm m}$, $s = {\rm mm}$, and $\beta = 0.5$, we find that the ratio 
between the growth time and the drift time $t_{\rm drift} = r / |v_r|$ is,
\begin{equation}
 \frac{t_{\rm grow}}{t_{\rm drift}} \approx 87 
 \left( \frac{\alpha}{10^{-2}} \right)^{-1} 
 \left( \frac{h/r}{0.05} \right)^2 
 \left( \frac{\dot{M} / \dot{M}_p}{10} \right) 
 \tau_s^2.
\end{equation} 
For these parameters, the track given by equation~(\ref{eq_track}) intersects the allowed 
streaming region at $\tau_s \simeq 0.035$, at which point the growth time is a small 
fraction of the drift time. This ordering is not, however, robust, and for larger $\tau_s$ 
growth may be slower than drift.

\begin{figure}
\begin{center}
\includegraphics[width=0.45\textwidth,angle=0]{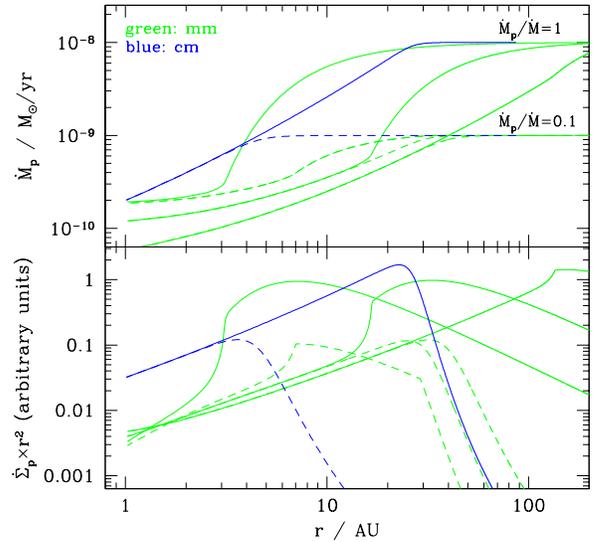}
\end{center}
\vspace{-0.3truein}
\caption{The radial projection of the models shown in Figure~\ref{fig1}. Upper panel: the radial 
dependence of the particle mass accretion rate. Lower panel: the radial dependence of 
planetesimal formation. Solid curves refer to models with $\dot{M}_p / \dot{M} = 1$, 
dashed curves models with $\dot{M}_p / \dot{M} = 0.1$. The assumed planetesimal formation 
time for the mm-sized particles cases increases from right to left ($K=10^2, 10^3, 10^4$). 
For cm-sized particles only the case with $K=10^2$ is plotted.}
\label{fig2}
\end{figure}

To model the impact of planetesimal formation on the track in $Z$-$\tau_s$ space, we 
integrate the continuity equation for the particles,
\begin{equation}
 \frac{\partial \Sigma_p}{\partial t} + \frac{1}{r} \frac{\partial}{\partial r} \left( r \Sigma_p v_r \right) = 
 - \frac{\Sigma_p}{t_{\rm form}},
\end{equation}
to find a steady-state solution. 
We assume that planetesimals form on a multiple $K$ of the dynamical time scale 
within the allowed region, $t_{\rm form} = K \Omega^{-1}$. For $Z < Z_{\rm crit}$ 
we smoothly suppress the rate by a factor $\exp [ (Z_{\rm crit}-Z)/0.1 Z_{\rm crit} ]$. 
This roll-off in the planetesimal formation rate is imposed for numerical 
convenience, though it is physically plausible that some planetesimal formation 
persists at metallicities just below the nominal threshold. We consider 
icy particles ($\rho_m = 1 \ {\rm g \ cm}^{-3}$) in a disk with $\dot{M} = 10^{-8} \ M_\odot \ {\rm yr}^{-1}$, 
$\alpha = 10^{-2}$, and $T = 150 ( r / 3 \ {\rm AU})^{-0.5} \ {\rm K}$.

Figure~\ref{fig1} shows solutions to this model. The key point is that, within the 
allowed region, the time scale for prompt planetesimal formation scales with radius as 
$r^{-3/2}$, which can be compared to the time scale for radial drift  
$t_{\rm drift} = {\rm const}$. There is therefore a critical radius within which 
planetesimal formation dominates, whereas outside radial drift leads to particle pile-up 
\citep{youdin04}. If the streaming instability leads to gravitational collapse on an essentially 
dynamical time scale \citep[$K \sim10^2$, as is found in simulations, e.g.][]{simon16}, then 
the derived $Z$-$\tau_s$ tracks skirt the lower boundary of the allowed region. Slower 
planetesimal formation time scales lead to clearly defined regions where first radial drift and 
then planetesimal formation dominate. We remark that models with large values of $K$ have an alternate 
interpretation in terms of stochastic planetesimal formation. For example, a model with 
$\dot{M_p} / \dot{M} = 1$ and $K=10^4$ is equivalent to one with $\dot{M_p} / \dot{M} = 10^{-2}$ 
if transient local concentrations attain the metallicity needed for prompt planetesimal formation 
(with $K=10^2$) 1\% of the time.

Figure~\ref{fig2} shows the solutions as a function of radius (notionally extended in to 1~AU, though 
this would be inward of the physical snow line). For $\dot{M}_p / \dot{M} \gtrsim 0.1$ and mm- to cm-sized 
particles, planetesimal formation in the outer disk is an efficient sink for the radial 
flow of solids. The mass accretion rate of solids reaching the snow line is substantially reduced, by a factor 
that ranges from several to more than an order of magnitude. As a result, the rate of planetesimal 
formation at any given radius depends on whether planetesimals have formed further out in the disk. For most 
of the models considered the sweet spot for planetesimal formation is well outside the snow line, 
in some cases as far out as 30-40~AU. Planetesimals are 
predicted to be laid down across a broad range of radii for high particle fluxes, whereas lower 
fluxes yield a narrower distribution. The outer cut-off will be sharp if, as simulations 
suggest \citep{johansen09}, planetesimal 
formation becomes rapidly inefficient below a threshold metallicity.

\section{Particle and vapor diffusion}
We now explore how robust these conclusions are to neglected physical effects. 
Turbulent particle diffusion works against particle pile-up \citep{hughes12}, further 
favoring large radii as sites for planetesimal formation, but the condensation of 
vapor diffusing across the snow line has the opposite effect \citep{stevenson88}.
We adopt a time-dependent treatment \citep{alexander07} and solve continuity equations that treat 
solids and vapor as trace species, with surface density $\Sigma_t$,
\begin{equation}
 \frac{\partial \Sigma_t}{\partial t} + \frac{1}{r} \frac{\partial}{\partial r} \left[ r \left( F_{\rm diff} + \Sigma_t v_r \right) \right] = 0.
\end{equation} 
The diffusive term $F_{\rm diff} = -D \Sigma \partial (\Sigma_t / \Sigma)/ \partial r$ \citep{clarke88}, where $D$ is the diffusion 
co-efficient (here set equal to $\nu$ for both particles and vapor), and $v_r$ is the gas radial velocity 
(for the vapor) or the aerodynamic drift speed (for particles). We supplement these equations with 
source / sink terms appropriate for instantaneous sublimation and condensation of water vapor, 
following the method and chemical constants given in \cite{ciesla06}. In brief, at radii where ice is not 
fully sublimated, we sublimate or condense the appropriate amount of ice or vapor at each time step 
to maintain the actual vapor pressure of water at the equilibrium value given by the Clausius-Clapeyron 
equation. We assume that there is no significant change to the size distribution of particles in the 
vicinity of the snow line , an assumption which minimizes its importance. In more complete 
models, at least some vapor condenses on to pre-existing particles, growing them to larger sizes 
\citep{ros13}.

\begin{figure}
\begin{center}
\includegraphics[width=0.45\textwidth,angle=0]{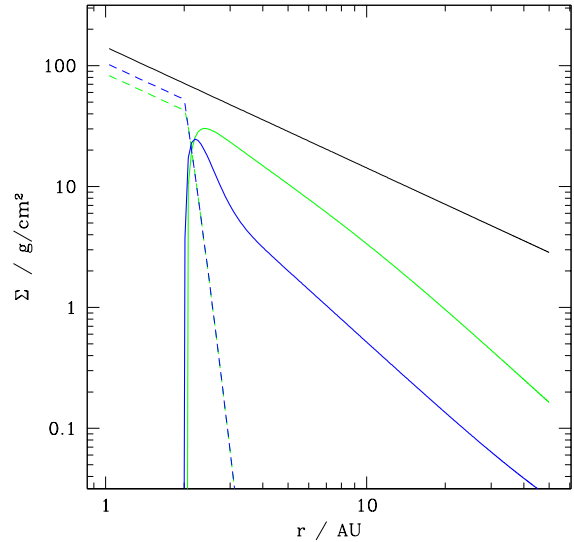}
\end{center}
\vspace{-0.3truein}
\caption{Example steady-state surface density profiles of mm-sized (green) and cm-sized (blue) particles, 
water vapor (dashed lines), and gas, for the disk model with $\dot{M} = 10^{-8} \ M_\odot \ {\rm yr}^{-1}$. 
Outward diffusion and condensation of vapor leads to a modest enhancement of the solid surface density 
outside the snow line for cm-sized (and larger) particles.}
\label{fig3}
\end{figure}

Figure~\ref{fig3} shows the resulting steady-state profiles of vapor and solids, computed in the 
fiducial disk model. The solid metallicity (which is freely scalable in models that ignore planetesimal 
formation) is set to match the analytic results for $\dot{M_p} / \dot{M} = 1$ at 50~AU. For 
$r \gtrsim 5 \ {\rm AU}$ the surface density of solids is close to a power-law, consistent with 
turbulent diffusion being a minor effect on these scales. For mm-sized particles the condensation 
of diffusing vapor also makes a negligible change to the equilibrium distribution of solids, but for 
cm-sized (and larger) solids a modest enhancement of solids upstream of the snow line occurs. 
Slightly stronger effects are possible if $D / \nu > 1$. Because the radial drift velocity exceeds the 
gas velocity, the abundance of water vapor interior to the snow line is enhanced as long as 
large masses of small solids remain present in the disk.

\begin{figure}
\begin{center}
\includegraphics[width=0.45\textwidth,angle=0]{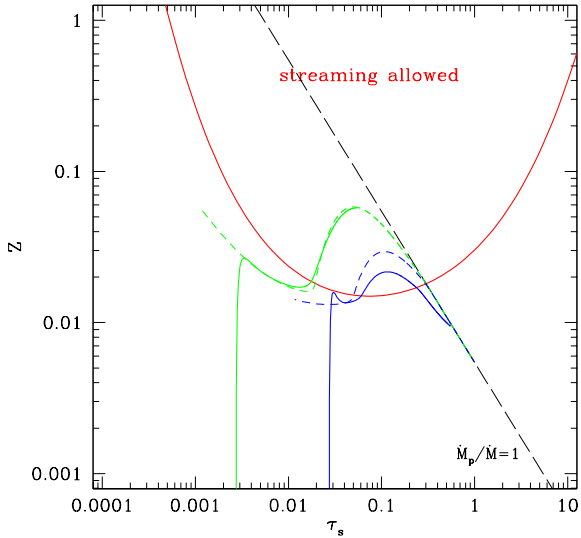}
\end{center}
\vspace{-0.3truein}
\caption{Steady-state tracks for mm-sized (green curves) and cm-sized (blue) particles in $Z$-$\tau_s$ 
space, including the effects of turbulent particle diffusion and vapor diffusion / condensation. We assume 
$\dot{M}_p / \dot{M} = 1$, $K=10^3$, and the fiducial disk model. The simpler 
models shown in Figure~\ref{fig1} are plotted as the dashed lines.}
\label{fig4}
\end{figure}

Adopting the same ``scale-free" model for local planetesimal formation losses as in \S\ref{sec_analytic}, 
we show in Figure~\ref{fig4}  an illustrative example of how the results shown in Figure~\ref{fig1} are modified by turbulent 
diffusion and vapor condensation. As before, the gas disk has $\dot{M} = 10^{-8} \ M_\odot 
\ {\rm yr}^{-1}$ and $T = 150 ( r / 3 \ {\rm AU})^{-0.5} \ {\rm K}$, which places the snow line at 
radii similar to those inferred for the Solar System \citep{morbidelli00}. As was already 
obvious from Figure~\ref{fig3}, for mm-sized particles the extra physics 
in the numerical model makes very little difference to the predicted radii where planetesimals  
could form. For cm-sized particles there is a more significant deviation from the analytic 
model results. The higher surface density caused by the condensation 
of diffusing water vapor results in a secondary peak of planetesimal formation just outside 
the snow line, but does not alter the conclusion that the most-favored location lies 
further out. This result would be reinforced at higher {\em gas} 
accretion rates, arguably more appropriate to an early phase of disk evolution, which 
would boost the threshold size for vapor condensation effects to matter.

\section{Discussion}
At radii beyond the water snow line there is a limited window to prompt planetesimal 
formation that invokes only known physical processes: coagulation to a fixed size that is of the 
order of mm, radial drift, 
and gravitational collapse of streaming-initiated over-densities. Planetesimal formation via this route 
is possible early on --- while radial drift is rapidly reducing the global dust-to-gas ratio --- and would 
typically lead to the deposition of a broad and massive belt of planetesimals well outside the snow 
line. We have not attempted fine tuning of the model, but Figure~\ref{fig2} makes it 
clear that features such as the outer edge to the Kuiper Belt \citep{trujillo01} and 
the large masses of primordial debris required in the Nice model \citep{tsiganis05}, are 
qualitatively consistent with planetesimal formation expectations, as noted by 
\citet{youdin02}.

Efficient early planetesimal formation is a prerequisite for early planet formation, which 
has been proposed as an explanation for the ring-like structures seen in ALMA observations of 
HL~Tau \citep{hltau15}. There is tension, however, between other observations and {\em any} 
model that invokes radial drift and pile-up as key ingredients of planetesimal formation. Such 
models predict a rapid loss of small solids via radial drift and planetesimal formation, whereas 
observations of CO line emission toward mostly older sources favor dust-to-gas ratios in 
{\em excess} of $10^{-2}$ \citep{williams14,eisner16}. A better understanding of gas disk 
mass estimates, which for now differ substantially depending on the technique used 
\citep[e.g.][]{bergin13,manara16}, is the single most important advance that would constrain 
theoretical models of radial drift.

Our model is deliberately simple, and is intended to provide insight into the 
physical origin and robustness of more complex approaches. It could be improved by better 
delineating the conditions needed for planetesimal formation \citep[e.g. by including the 
dependence on the radial pressure gradient,][]{bai10}, and by modeling the coupled growth and drift of 
particles. Several authors have already developed such coupled models. \cite{drazkowska14}, using a 
Monte Carlo dust coagulation code, argued for preferential triggering of the streaming instability 
beyond the ice line. Similarly, \cite{krijt16}, using a detailed model for particle growth 
coupled to estimates of planetesimal formation thresholds similar to ours, found that 
dust-rich disks with weak turbulence promote prompt outer disk planetesimal formation. 
In the inner disk, there is general agreement that planetesimal formation is harder, and 
a prompt route is only possible if particles grow to substantially larger sizes 
\citep[limited only by fragmentation and radial drift, rather than bouncing,][]{drazkowska16}. 
Alternatively, planetesimal formation in the terrestrial planet region may occur via a less-efficient 
stochastic channel \citep[driven by local turbulent enhancements in $Z$,][]{johansen14}, and at 
still smaller radii at persistent traps associated with the inner edge of the dead zone \citep{lyra09}. 
Despite their differences all models agree that attaining the conditions needed for planetesimal 
formation is relatively hard, and that as a result the initial distribution of 
planetesimals is unlikely to be a simple power-law but rather a complex function of radius.

Building the cores of the giant planets requires, first, the deposition of a large mass of 
solids into planetesimals \citep{pollack96}. The preference of the streaming instability for relatively 
large stopping times 
challenges the common view that the favored location for this deposition is adjacent 
to the snow line. Even when the effects of vapor diffusion and condensation are 
considered, we find that the best time and place to lay down a large mass of planetesimals is 
early and at large radii. This suggests a model for core formation in which planetesimal 
formation and initial growth occurs well beyond the snow line. Subsequent growth 
could either occur in situ via pebble accretion \citep{lambrechts14,levison15}, 
or via migration to a trap closer in.

\acknowledgments
We thank the referee for a detailed and instructive report. 
PJA thanks Cathie Clarke and the Institute of Astronomy, Cambridge, for hospitality, and 
acknowledges support from NASA through grants NNX13AI58G and NNX16AB42G, and from NSF 
AAG grant AST 1313021. JAE acknowledges support from NSF AAG grant 1211329. JBS's support was 
provided in part under contract with
the California Institute of Technology (Caltech) and the Jet Propulsion
Laboratory (JPL) funded by NASA through the Sagan Fellowship Program
executed by the
NASA Exoplanet Science Institute.

\bibliographystyle{aasjournal}

\end{document}